\documentclass{emulateapj}

\usepackage{savesym}
\savesymbol{tablenum}
\usepackage{siunitx}
\restoresymbol{SIX}{tablenum}

\let\oldendlongtable\endlongtable
\let\oldlongtable\longtable
\makeatletter
\renewenvironment{longtable}[1]{
\oldlongtable{[#1]}
\def\pt@format{#1}
}{\oldendlongtable}
\makeatother

\usepackage{color}
\definecolor{white}{rgb}{1,1,1}

\usepackage{colortbl}
\definecolor{lmugreen}{rgb}{0.01,0.58,0.25}

\usepackage{amsmath}
\usepackage{natbib}
\usepackage{float}
\usepackage{dcolumn}
\usepackage{booktabs}
\usepackage[
	colorlinks=true,
	urlcolor=lmugreen,
	linkcolor=lmugreen,
	citecolor=lmugreen
	]{hyperref}
\usepackage{rotating}
\usepackage{graphicx}
\usepackage{tablefootnote}
\usepackage{longtable}
\usepackage[nolist]{acronym}

\usepackage{listings}
\newcommand{\citeorder}[1]{}

\defcitealias{KDR1:inprep}{Paper I}
\defcitealias{KDR2a:inprep}{Paper II}
\defcitealias{KDR2b:inprep}{Paper III}
\defcitealias{KDR3:inprep}{Paper IV}
\defcitealias{KoepferlRobitaille:inprep}{Paper FluxCompensator}

\DeclareSIUnit\yr{yr}
\DeclareSIUnit\ergs{ergs}
\DeclareSIUnit\Jy{Jy}
\DeclareSIUnit\Hz{Hz}
\DeclareSIUnit\sr{sr}
\DeclareSIUnit\pc{pc}
\DeclareSIUnit\ppm{ppm}
\DeclareSIUnit\kpc{\kilo\pc}
\DeclareSIUnit\AU{AU}
\DeclareSIUnit\eqsign{\text{\ensuremath{=}}}
\DeclareSIUnit\simeqsign{\text{\ensuremath{\simeq}}}
\DeclareSIUnit\mag{mag}
\DeclareSIUnit\gramms{g}
\DeclareSIUnit\gccm{\gramms\per\cubic\centi\metre}
\DeclareSIUnit\microns{\micro\metre}
\DeclareSIUnit\Myr{\mega\yr}
\DeclareSIUnit\Msun{\text{\ensuremath{M_\odot}}}
\DeclareSIUnit\Lsun{\text{\ensuremath{L_\odot}}}
\DeclareSIUnit\Rsun{\text{\ensuremath{T_\odot}}}
\DeclareSIUnit\Minfall{\Msun\per\yr}

\slugcomment{The Astrophysical Journal}

\shorttitle{Synthetic Star-forming Regions - III. Calibration of Measurement Techniques of Star-formation Rates}
\shortauthors{Koepferl, Robitaille, Dale}

\begin{document}

\title{Insights from Synthetic Star-forming Regions: \\III. Calibration of Measurement Techniques of Star-formation Rates}
\author{Christine M. Koepferl$^{1,2}$, Thomas P. Robitaille$^{1,3}$, and James E. Dale$^4$}
\affil{$^1$ Max Planck Institute for Astronomy, K\"onigstuhl 17, D-69117 Heidelberg, Germany\\
$^2$ Scottish Universities Physics Alliance (SUPA), School of Physics and Astronomy, University of St Andrews\\North Haugh, St Andrews, KY16 9SS, UK\\
$^3$ Freelance Consultant, Headingley Enterprise and Arts Centre, Bennett Road Headingley, Leeds LS6 3HN\\
$^4$ University Observatory Munich, Scheinerstr. 1, D-81679 Munich, Germany}
\email{cmk8@st-andrews.ac.uk}
\received{29 January 2016}
\accepted{8 March 2016}

\begin{abstract}
Through an extensive set of realistic synthetic observations (produced in \citetalias{KDR1:inprep}), we assess in this part of the paper series \citepalias{KDR2b:inprep} how the choice of observational techniques affects the measurement of \acp{SFR} in star-forming regions. We test the accuracy of commonly used techniques and construct new methods to extract the \ac{SFR}, so that these findings can be applied to measure the \ac{SFR} in real regions throughout the Milky Way. We investigate diffuse infrared \ac{SFR} tracers such as those using \SI{24}{\microns}, \SI{70}{\microns} and total infrared emission, which have been previously calibrated for global galaxy scales. We set up a toy model of a galaxy and show that the infrared emission is consistent with the intrinsic \ac{SFR} using extra-galactic calibrated laws (although the consistency does not prove their reliability). For local scales, we show that these techniques produce completely unreliable results for single star-forming regions, which are governed by different characteristic timescales. We show how calibration of these techniques can be improved for single star-forming regions by adjusting the characteristic timescale and the scaling factor and give suggestions of new calibrations of the diffuse star-formation tracers. We show that star-forming regions that are dominated by high-mass stellar feedback experience a rapid drop in infrared emission once high-mass stellar feedback is turned on, which implies different characteristic timescales. Moreover, we explore the measured \acp{SFR} calculated directly from the observed young stellar population. We find that the measured point sources follow the evolutionary pace of star formation more directly than diffuse star-formation tracers.
\end{abstract}

\section{Introduction}
\label{C5:Sec:Introduction}
In a star-forming region, gas is transformed into stars. While the dominating driving mechanisms and the impact of the different stellar feedback mechanisms are still debated in the field, there are different approaches among researchers to address these questions: In the observational approach, the characteristic observable properties of star-forming regions are used to study and compare different regions in an attempt to disentangle the dominant driving mechanisms of star formation. In the theoretical approach, (magneto-)hydrodynamical simulations (e.\,g.~\acs{SPH}: \acl{SPH}) study the formation processes of star-forming regions. However, in state-of-the-art simulations of star-forming regions the total \ac{SFR} is always too high and some simulations fail to reproduce the correct distribution of stellar masses when comparing with observations (c.\,f.~\citealt{Krumholz:2012SPH2}, \citealt{Krumholz:2011}, \citealt{Urban:2010}, \citealt{Bate:2009} and also \citealt{DaleI:2011,DaleIoni:2012,DaleIoni:2013,DaleWind:2013,DaleBoth:2014}, referred to as \acs{D14} \acs{SPH} simulations). Different implementations of stellar feedback, such as ionization and winds, help to suppress star formation and the accretion mechanism, however, these mechanisms do not restrain star formation enough (c.\,f.~\citealt{Bate1:2014}, \citealt{Federrath:2014}, \citealt{Hansen:2012}, and also \acs{D14} \acs{SPH} simulations). For more detail about high-mass stellar feedback and its implementation, see the review of \cite{Dale:Review:2015}.  

\newpage
The measured star-formation properties are the observational benchmark between reality and theory. Examples of properties include the total gas mass $M_{\textup{gas}}$ and the total stellar mass $M_*$ in a star-forming region \citep[as summarized by][]{DaleBoth:2014}. The property which connects stellar mass and total gas mass is the efficiency of the transition from gas to stars and can be expressed by the \ac{SFE} at a certain time $t$:
\begin{eqnarray}
    \textup{SFE}(t)&=&\frac{M_*(t)}{M_*(t) + M_{\textup{gas}}(t)}.
\end{eqnarray}
Another important physical property defining a star-forming region is the rate at which it forms stars. The average \ac{SFR} between times $t_0$ and $t$ is defined as: 
\begin{eqnarray}
    \label{C5:Eq:SFE}
    \textup{SFR}(t)&=&\frac{M_*(t)-M_*(t_0)}{t-t_0} = \frac{M_*(t)-M_*(t_0)}{\delta t_*},
\end{eqnarray}
where $\delta t_* =t-t_0$ is the characteristic timescale over which the \ac{SFR} is measured. There exists a large variety of tracers to measure the \ac{SFR}. Each technique probes star formation within a certain characteristic timescale $\delta t_*$, which varies from tracer to tracer. One can divide the different techniques into two categories: 
\begin{enumerate}
\item {\bf Indirect Techniques}\\Young high-mass stars ionize the material around them with X-ray and \ac{UV} radiation. This process can be used to trace star formation in different wavelength regimes, since the ionized electrons emit free-free centimeter continuum, while ionized hydrogen emits H$\alpha$ radiation in a bubble around the young ionizing stars. The ionization strength is correlated with the mass of the ionizing objects \citep[e.\,g.][]{Panagia:1973}. Moreover, young high-mass objects also emit weaker \ac{UV} radiation which is then absorbed by the surrounding dust and produces diffuse infrared emission. However, in this picture, the thermal emission of young low-mass stars is neglected. Consequently, these methods, which make use of the free-free, H$\alpha$ and diffuse infrared emission (originating from \ac{UV}), are indirect methods, since they do not trace the actual young stars but the effect of the high-mass young stars on their surrounding dust and gas. Since these indirect techniques only trace the high-mass stars, the findings need to be extrapolated to the total stellar population by assuming an \ac{IMF}. For more details on indirect tracers, see  the review of \cite{KenniEvans:2012} and \cite{Calzetti:2013}.
\item {\bf Direct Techniques}\\Direct star-formation tracers trace the sites of young stars directly. Young stellar objects (\acsp{YSO}) heat the surrounding dust which then emits in the infrared. In the circumstellar material, the dominating \SI{24}{\microns} emission originates closer to the star than the \SI{70}{\microns} emission. The \SI{24}{\microns} emission can act as a good tracer since the resolution is currently better than for \ac{FIR} emission. Moreover, the \SI{24}{\microns} emission is unrelated to the large scale infrared emission from hydrodynamical temperature fluctuations when comparing to the \ac{FIR} emission which is background dominated (\citealt{KDR2a:inprep}, referred to as \citetalias{KDR2a:inprep}). However, selecting a sample of young objects is not a trivial task, as shown by \cite{Koepferl:2015} for the \ac{CMZ}.
\end{enumerate}
The measurement of star-formation tracers, such as the \ac{SFR}, requires different approaches for different astronomical objects, such as galaxies, giant molecular clouds or single star-forming regions: 
\begin{enumerate}
\item {\bf Extra-galactic Scales}\\For distant galaxies, individual counting of forming stars is not possible, since not even single star-forming regions are resolved. The measured properties of star formation are therefore averaged over large scales. Hence, indirect, atomic, recombination line tracers (e.\,g.~H$\alpha$) or indirect diffuse tracers (e.\,g.~\SI{24}{\microns}, \SI{70}{\microns}, total infrared) are commonly used in the extra-galactic community. Those indirect methods have been calibrated using galaxy model spectra, such as STARBURST99 \citep{Leitherer:1999}, and are constantly improved empirically (e.\,g.~\citealt{Murphy:2011, Calzetti:2010, Rieke:2009,Pannella:2015,Leroy:2013,Leroy:2012,Wuyts:2011,Leroy:2008}).
\item {\bf Local Galactic Scales}\\For local regions within the Milky Way, the \ac{SFR} can be directly evaluated through proto-stellar counting. Using young stars directly has the clear advantage that the sites of star formation are counted and not the total emission which is produced by star formation or other processes. For example, \cite{Evans:2009} estimated the \ac{SFR} for regions in the Milky Way through directly counting \acsp{YSO} in the \ac{MIR} (from the \ac{c2d} project) down to low-mass objects. For their complete sample, they assumed an average mass of \SI{0.5}{\Msun} and a characteristic timescale of $\delta t_*=\SI{2}{\Myr}$. \cite{Robitaille:2010} estimated the \ac{SFR} of the Milky Way by directly counting \acsp{YSO} emitting in the infrared (\acl{GLIMPSE}, \acs{GLIMPSE}), using a population synthesis model to extrapolate the number of sources below the detection limit. In addition, indirect tracers, such as free-free emission, have been applied for sub-regions of the Milky Way. For example, \cite{Longmore:2013} measured the \ac{SFR} of the \ac{CMZ} using free-free emission. Diffuse infrared tracers (e.\,g.~\SI{24}{\microns}, \SI{70}{\microns}, total infrared) have also been used for the Milky Way and single star-forming regions therein \citep{Barnes:inprep,Crocker:2011,Yusef:2009,Misiriotis:2006}. However, indirect tracers have not yet been critically tested for the Milky Way.
\item {\bf Intermediate Scales}\\
Keeping in mind that indirect methods have not yet been calibrated for regions within the Milky Way, it is challenging to measure the \ac{SFR} for regions across the Galactic plane where the detection of point sources is limited. One of the major flaws of the star-formation tracers is that they have been calibrated from far-to-near. However, with better infrared observations (Galactic and extra-galactic) at hand and state-of-the-art radiative transfer codes, we now have the opportunity to improve the relation between the star-formation tracers and the \ac{SFR} from the bottom up. Once reliable direct techniques to estimate the star-formation properties have been calibrated at local scales, they can be used to calibrate indirect techniques for local regions, which then also can be used for regions across the Galaxy and for distant galaxies.
\end{enumerate}

When calibrating different techniques to infer star-formation properties, the discussion of the timescales $\delta t_*$ is an essential ingredient. Note that different techniques, which make use of different physical processes that act on different timescales $\delta t_*$, produce different \acp{SFR} for the same region if the intrinsic rate is not constant over time. On small scales, the \ac{SFR} is not necessarily expected to be constant. On the other hand, on the scale of the Galaxy, the rate of star formation is expected to be approximately constant on $\sim$\,\SI{100}{\Myr} timescales and hence the different techniques should produce comparable rates. For example, \cite{Chomiuk:2011} showed that the measured \ac{SFR} values for the Milky Way \citep[e.\,g.~][]{Robitaille:2010,Murray:2010,Misiriotis:2006} agreed when scaling by timescale to a value of \SI{1.9(4)}{\Minfall}.

\newpage
\subsection{Motivation}
Since the observable star-formation properties are used to test theoretical simulations, which might then help us to understand the main driving mechanisms of star formation, it is essential that the measurements of these properties are as accurate as possible. In this work, we test the accuracy and calibrate relations between star-formation tracers and physical properties, such as the total gas mass $M_{\textup{gas}}$, the \ac{SFR} and the total stellar mass $M_*$, using indirect and direct techniques. In \cite{KDR2a:inprep}, referred to as \citetalias{KDR2a:inprep}, we tested the accuracy of measurements of the total gas mass. In this part of the series, referred to as \citetalias{KDR2b:inprep}, we will focus on techniques that make use of dust tracers to infer star-formation rates, using the set of realistic synthetic observations developed in \cite{KDR1:inprep} referred to as \citetalias{KDR1:inprep}.

\subsection{Recapitulation}
\label{C5:Sec:recap}
In the following we will briefly summarize \citetalias{KDR1:inprep}, where we produced synthetic observations of multiple time-steps of a simulated star-forming region. We used the \acs{D14} \acs{SPH} simulations (\citealt{DaleI:2011,DaleIoni:2012,DaleIoni:2013,DaleWind:2013,DaleBoth:2014}) of a \SI{30}{\pc} wide synthetic star-forming region with different high-mass stellar feedback mechanisms implemented (\textit{run I}: stellar ionization and winds). The configurations of the simulated star-forming from the \acs{D14} \acs{SPH} simulations are described in more detail in \citetalias{KDR1:inprep} and \acs{D14}.

We extend the \acs{SPH} simulations by post-processing the simulation output through radiative transfer calculations. We account for the stellar heating of the dust using \textsc{Hyperion}, a 3-d dust continuum Monte-Carlo radiative transfer code. For more details about \textsc{Hyperion} and radiative transfer in general, see \cite{Robitaille:2011} and \cite{Steinacker:2013}.

In \citetalias{KDR1:inprep}, we presented a mass conserving algorithm which converts an \acs{SPH} particle distributions (e.\,g.~density) to a Voronoi tessellation used by the radiative transfer code. In order to better recover the \ac{MIR} flux, we refined the density structure close to \acsp{YSO} to overcome the resolution limit of the \acs{SPH} simulation. We tested different extrapolation techniques for the refinement: rotationally flattened \citep{Ulrich:1976} envelope profile and a power-law envelope profile. In order to improve the computational efficiency, we precomputed with the radiative transfer code analytical models of \acsp{YSO} within \SI{500}{\AU} of every accreting proto-star. The typical dust-to-gas ratio of $\frac{\textup{dust}}{\textup{gas}}=\num{0.01}$ \citep{DraineBook} was used for the radiative transfer calculations together with the \cite{Draine:2007} \ac{PAH} dust grains description.
Moreover, we combined the calculated radiative transfer dust temperature with an ambient background temperature typical for the relatively cloudless regions in the Galactic plane ($T_{\textup{iso}}=\SI{18}{\kelvin}$, see also \citetalias{KDR2a:inprep}). For more information about the radiative transfer set-up of synthetic star-forming regions, see \citetalias{KDR1:inprep}.

We used the \textsc{FluxCompensator} to produce realistic synthetic observations for all the radiative transfer images, by accounting for the extinction, the transmission curves of the telescope and detector, the pixel-size and the \ac{PSF} convolution. 
We combined these realistic synthetic observations with a relatively empty patch of the Galactic plane (see also \citetalias{KDR2a:inprep}) resulting in synthetic observations which are directly comparable to real observations. For more information about the \textsc{FluxCompensator}, see \citeauthor[in preparation]{KoepferlRobitaille:inprep}, referred to as \citetalias{KoepferlRobitaille:inprep}. We present the produced $\sim\,$5800 realistic synthetic observations in the online material of \citetalias{KDR1:inprep} resulting from combinations of the following configurations:
\begin{itemize}
\item {\bf 23 Time-steps}\\
We select 23 equally spaced time-steps (step width: $\Delta t=\SI{149000}{\yr}$) over \SI{3.3}{\Myr} from the \acs{SPH} simulations between the formation of the first proto-star and the first supernova. Once three high-mass stellar particles above \SI{20}{\Msun} have formed ($\sim$\,\SI{1.7}{\Myr}) the ionization and winds of high-mass stars are switched on.
\item {\bf 3 Circumstellar Set-ups}\\
We explored three different circumstellar refinement scenarios: as a control run one without added envelopes (\acs{CM1}) and two envelope refinements beyond the resolution limit of the simulation. In the two other cases we used pre-computed analytical models of a circumstellar disk together with a rotationally flattened envelope \citep[\acs{CM2},][]{Ulrich:1976} and a power-law envelope (\acs{CM3}) profile.
\item {\bf 3 Orientations}\\
We constructed synthetic observations for three perpendicular viewing angles: \ac{O1}, \ac{O2} and \ac{O3}.
\item {\bf 2 Distances}\\
We pushed the synthetic observations to two different distances: \SI{3}{\kpc} (\acs{D1}) and \SI{10}{\kpc} (\acs{D2}). We chose this distances in order to be comparable to 
nearby high-mass star-forming regions, such as Carina, Westerhout 4,5 and the Eagle Nebula and more distant regions across the Galactic plane. Using a built-in function of the \textsc{FluxCompensator}, we calculate the interstellar extinction \citep{Kim:1994} with $A_V=\num{10}$ and $A_V=\num{20}$ for the two different distances, respectively.
\item {\bf 7 Bands}\\
We produced realistic synthetic observations in the following bands with the respective appropriate transmission curve, \ac{PSF} and pixel-size (listed in brackets): \ac{IRAC} \SI{8}{\microns} (\ang{;;1.2}), \ac{MIPS} \SI{24}{\microns} (\ang{;;2.4}), \ac{PACS} \SI{70}{\microns} (\ang{;;3.2}), \ac{PACS} \SI{160}{\microns} (\ang{;;4.5}), \ac{SPIRE} \SI{250}{\microns} (\ang{;;6.0}), \ac{SPIRE} \SI{350}{\microns} (\ang{;;8.0}) and \ac{SPIRE} \SI{500}{\microns} (\ang{;;11.5}).
\item {\bf 2 Backgrounds}\\
We analyze every realistic synthetic observations and also a respective counterpart combined with a realistic background.
\end{itemize}
For more information about the derivation of realistic synthetic observations, see \citetalias{KDR1:inprep} and \citetalias{KoepferlRobitaille:inprep}.

\subsection{Outline}
With these realistic synthetic observations, which are directly comparable to real observations, we can start our analysis of star-formation dust tracers. The reliability of three different diffuse star-formation dust tracers will be explored on local scales in Section~\ref{C5:Sec:dSFR}. In Section~\ref{C5:Sec:dSFR_Time-scales} we calibrate these relations with different characteristic timescales and we set constrains for these star-formation tracers on global scales. We provide a new indirect relation between the luminosity and the stellar mass in Section~\ref{C5:Sec:dSFRemission}. In Section~\ref{C5:Sec:cSFR}, we present insights of the proto-stellar counting of our synthetic point sources. In Section~\ref{C5:Sec:Discussion}, we discuss the biases of star-formation properties before we summarize in Section~\ref{C5:Sec:Summary}. In a follow-up paper (\citeauthor[][in preparation]{KDR3:inprep}, referred to as \citetalias{KDR3:inprep}), we will present a detailed parameter study of our synthetic point-source catalogues and the resulting \ac{SFR}.

\begin{figure*}[t]
    \includegraphics[width=\textwidth]{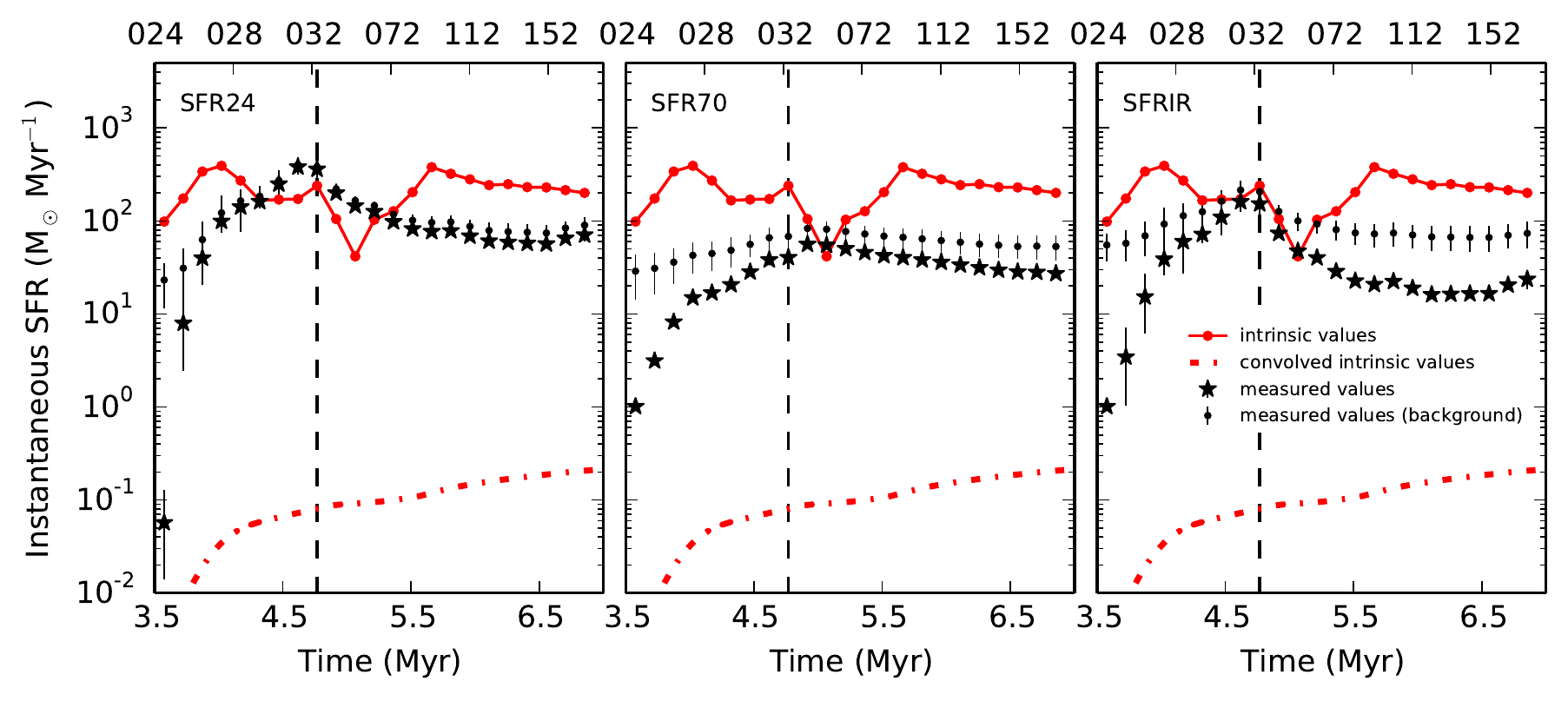}
  \caption[Averaged SFRs measured at \SI{24}{\microns}, \SI{70}{\microns} and the total infrared]{\label{C5:Fig:dSFR}Averaged measured \acp{SFR} with the techniques \acs{SFR24}, \acs{SFR70} and \acs{SFRIR} using the total flux at \SI{24}{\microns}, \SI{70}{\microns} and the total infrared flux, respectively, in comparison with the intrinsic \ac{SFR}. }
\end{figure*}
%
\section{Diffuse Star-formation Tracers}
\label{C5:Sec:dSFR}
There exist a variety of indirect diffuse infrared \ac{SFR} tracers which use the total flux in certain bands to derive the \ac{SFR}. For a detailed review, see \cite{KenniEvans:2012} or \cite{Calzetti:2013} or Section~\ref{C5:Sec:Introduction}. These techniques, which use tracers in the infrared continuum, rely on the assumption that young high-mass stars emit strongly at \ac{UV} wavelengths, which is then completely absorbed by the dust, and then completely re-emitted in the infrared. Diffuse infrared star-formation tracers assume further that the flux originates from all high-mass stars formed since a certain timescale $\delta t_*$. The shortcoming here is that the radiation of the majority of smaller stars is neglected. 

The diffuse techniques use the total flux in a certain observed band with wavelength $\lambda$. The conversion from total flux, hence total luminosity $\nu L_\nu(\lambda)$, to \ac{SFR} always follows the same pattern (see \citealt{KenniEvans:2012}):
\begin{eqnarray}
    \label{SFRlaws}
    \frac{\textup{SFR}_\lambda}{\si{\Msun\per\yr}}&=&a \left(\frac{\nu L_\nu(\lambda)}{\si{\ergs\per\second}}\right)^b
\end{eqnarray}
In order to get the total luminosity $\nu L_\nu(\lambda)$ in a certain band (e.\,g.~\ac{MIPS} \SI{24}{\microns}), we calculate the total flux $\nu F_\nu(\lambda)$ of the respective realistic synthetic observations. For this and upcoming photometric calculations in this paper, we use the \textsc{FluxCompensator}; for a more detailed description, see \citetalias{KoepferlRobitaille:inprep}. Following \citetalias{KDR2a:inprep}, we correct for the background emission and deredden with the same optical extinction $A_V$ as before. The resulting corrected total flux $\nu F^{\textup{corr}}_\nu(\lambda)$ is in units of \si{\ergs\per\second\per\centi\meter\squared}. In the respective band, the total luminosity can then be estimated with the measured distance $D$, where we assume again the correct distance, as in \citetalias{KDR2a:inprep} and set in \citetalias{KDR1:inprep}.
\begin{eqnarray}
   \nu L_\nu(\lambda)&=&4 \pi D^2 \nu F^{\textup{corr}}_\nu(\lambda) 
\end{eqnarray}
While there are many relations to infer the \ac{SFR} from diffuse tracers, in what follows, we will limit ourselves to only test the accuracy of the techniques\footnote{In this paper, we tested the techniques developed by \cite{Murphy:2011}, \cite{Calzetti:2010} and \cite{Rieke:2009} to infer the \ac{SFR} from the infrared flux. However, there exist a vast variety of similar techniques \& calibrations \citep[c.\,f.~][]{Pannella:2015, Leroy:2013,Leroy:2012,Leroy:2008,Wuyts:2011}.} using dust tracers, which have been summarized in the review of \cite{KenniEvans:2012} and are commonly used in the community \citep[e.\,g.~][]{Barnes:inprep,Crocker:2011,Yusef:2009}. Note that these techniques have been developed by and for the extra-galactic community, hence for entire galaxies, and assume characteristic timescales of $\delta t_* = \SI{100}{\Myr}$ \citep{KenniEvans:2012,Murphy:2011,Calzetti:2010,Rieke:2009}. These techniques have not yet been tested sufficiently for (smaller, local) individual regions. Therefore, we analyze the reliability of these techniques in this section. 

In Figure~\ref{C5:Fig:dSFR}, we plot the intrinsic instantaneous \ac{SFR} directly from the simulation (red solid curve), which we call $\textup{SFR}_{\textup{sim}}(\delta t_* =\Delta t)$ from now on, since it was calculated with the star-formation timescale $\delta t_*$ equal to the time-step of the simulation $\Delta t$. To compare it to the measured techniques, which follow characteristic timescales of $\delta t_* = \SI{100}{\Myr}$, we convolve \citep[see][]{NumericalRecipes} $\textup{SFR}_{\textup{sim}}(\delta t_* =\Delta t)$ with a normalized top-hat function \citep[see][]{Bronstein} of width \SI{100}{\Myr}\footnote{The top-hat function is not symmetric and evaluates at every point the average over the last \SI{100}{\Myr}.}. The resulting convolved function (red dashed curve), hereafter called $\hat{\textup{SFR}}_{\textup{sim}}(\delta t_* =\SI{100}{\Myr})$, is around three orders of magnitude lower since $\Delta t /\delta t_*  = \SI{149000}{\yr} /\SI{100}{\Myr} \approx \num{e-3}$.
\subsection{Method SFR24 | \SI{24}{\microns} Tracer}
\label{C5:Sec:dSFR_SFR24}
\cite{Rieke:2009} empirically calibrated a \ac{SFR}-to-luminosity relation of \SI{24}{\microns} emission:
\begin{eqnarray}
    \label{C5:Eq:SFR_Rieke}
    \frac{\textup{SFR}_{\SI{24}{\microns}}(\delta t_* =\SI{100}{\Myr})}{\si{\Msun\per\yr}}&=&\num{2.03e-43}\times\\
    &&\times\frac{\nu L_\nu(\SI{24}{\microns})}{\si{\ergs\per\second}},\nonumber
\end{eqnarray}
which they found to be valid within $\SI{2.3e42}{\ergs\per\second}\leq\nu L_\nu(\SI{24}{\microns})\leq \SI{5.0e43}{\ergs\per\second}$. Nevertheless, we will use this technique for synthetic observations with a total flux less than $\SI{e40}{\ergs\per\second}$ to test whether the \acs{SFR24} technique can be used for single star-forming regions. 

Using Eq.~\ref{C5:Eq:SFR_Rieke}, we derive the $\textup{SFR}_{\SI{24}{\microns}}(\delta t_* =\SI{100}{\Myr})$ for our set of synthetic observations in \ac{MIPS} \SI{24}{\microns} for every time-step. We present the results when combining with a background (black dots) and without background (black stars) in Figure~\ref{C5:Fig:dSFR}. Note that we averaged all values of different orientations, distances and circumstellar set-ups to one value per time-step because the total flux is very similar between all cases. We highlight the deviation, which is due to optically thick regions for different orientations and slight flux deviations, due to the different circumstellar set-ups, by black error-bars. 

In Figure~\ref{C5:Fig:dSFR} (left panel), we can see that the \acs{SFR24} technique produces similar results for the measurements of the synthetic observations with a combined background (black dots) and without a background (black stars). The fact that the $\textup{SFR}_{\SI{24}{\microns}}(\delta t_* =\SI{100}{\Myr})$ for the models including the background is higher is likely due to an incomplete subtraction of the non-uniform background. Nevertheless, the measured values are up to three orders of magnitude higher than the rate from the simulation averaged over a similar timescale $\hat{\textup{SFR}}_{\textup{sim}}(\delta t_* =\SI{100}{\Myr})$. However, they should match if the technique worked accurately. We provide the measured values in Table~\ref{C5:Appendix:data_CMxOxDx} of Appendix~\ref{C5:Appendix}.
\subsection{Method SFR70 | \SI{70}{\microns} Tracer}
\label{C5:Sec:dSFR_SFR70}
Based on \textit{Spitzer} data, \cite{Calzetti:2010} empirically calibrated the \ac{SFR} using the total luminosity in \SI{70}{\microns} above $\nu L_\nu(\SI{70}{\microns})\geq \SI{1.4e42}{\ergs\per\second}$:
\begin{eqnarray}
    \label{C5:Eq:SFR_Calzetti}
    \frac{\textup{SFR}_{\SI{70}{\microns}}(\delta t_* =\SI{100}{\Myr})}{\si{\Msun\per\yr}}&=&\num{5.88e-44}\times\\
    &&\times\frac{\nu L_\nu(\SI{70}{\microns})}{\si{\ergs\per\second}}.\nonumber
\end{eqnarray}
Note that here we will use \ac{PACS} \SI{70}{\microns} to estimate the $\textup{SFR}_{\SI{70}{\microns}}(\delta t_* =\SI{100}{\Myr})$ with Eq.~\ref{C5:Eq:SFR_Calzetti} and again test the \acs{SFR70} for luminosities below $\SI{e39}{\ergs\per\second}$. We show the results again in Figure~\ref{C5:Fig:dSFR} (black symbols in middle panel). We can see that the difference between the measurement with combined background (black dots) and the background-free measurements (black stars) is larger than for the method \acs{SFR24}. This is again due to the high background of the Galactic plane (as explained in \citetalias{KDR2a:inprep}) and a result of the non-uniform background during background correction. Again, similarly to what we found for the \acs{SFR24} technique, the measured $\textup{SFR}_{\SI{70}{\microns}}(\delta t_* =\SI{100}{\Myr})$ is two orders of magnitude higher than the expected value from the simulation $\hat{\textup{SFR}}_{\textup{sim}}(\delta t_* =\SI{100}{\Myr})$. We provide the measured values in Table~\ref{C5:Appendix:data_CMxOxDx} of Appendix~\ref{C5:Appendix}.
\subsection{Method SFRIR | Total Infrared Tracer}
\label{C5:Sec:dSFR_SFRIR}
\cite{Murphy:2011} derived their star-formation relation from the STARBURST99 models \citep{Leitherer:1999} for the total infrared luminosity reaching from \SI{8}{\microns} to \SI{1000}{\microns}. 
\begin{eqnarray}
    \label{C5:Eq:SFR_Murphy}
    \frac{\textup{SFR}_{\textup{IR}}(\delta t_* =\SI{100}{\Myr})}{\si{\Msun\per\yr}}&=&\num{3.88e-44}\frac{L(\textup{IR})}{\si{\ergs\per\second}}.
\end{eqnarray}
We derive the total luminosity in the infrared under the assumption that our 7 bands from $\nu_0$ (\ac{IRAC} \SI{8}{\microns}) to $\nu_1$ (\ac{SPIRE} \SI{500}{\microns}) cover most of the infrared emission. We integrate (using Simpson's rule, see \citealt{Bronstein}, \citealt{NumericalRecipes}) to calculate the flux density $F^{\textup{corr}}_\nu(\lambda)$:
\begin{eqnarray}
    \label{C5:Eq:Simpson}
    L(\textup{IR})&=& \int\limits_{\nu_0}^{\nu_1}d\nu F^{\textup{corr}}_\nu(\lambda)
\end{eqnarray}
With the total infrared luminosity $L(\textup{IR})$, we use Eq.~\ref{C5:Eq:SFR_Murphy} to derive the $\textup{SFR}_{\textup{IR}}(\delta t_* =\SI{100}{\Myr})$ for our set of synthetic observations. Again, in Figure~\ref{C5:Fig:dSFR} (right panel), we compare the measurements (black) with the values from the simulations (red). Again the measurements with combined background (black dots) are higher due to the complex background. The offset is comparable to the technique \acs{SFR70} because \emph{Herschel} bands with high background have been used (see \citetalias{KDR2a:inprep}). For the background-free measurements (black stars), the simulation spread is again larger, which is due to the \ac{MIR} parts of the total infrared flux (see \acs{SFR24} and \acs{SFR70}). Overall, the technique \acs{SFRIR} produces values of the $\textup{SFR}_{\textup{IR}}(\delta t_* =\SI{100}{\Myr})$ which are up to three orders of magnitude higher than the simulated rate $\hat{\textup{SFR}}_{\textup{sim}}(\delta t_* =\SI{100}{\Myr})$. We provide the measured values in Table~\ref{C5:Appendix:data_CMxOxDx} of Appendix~\ref{C5:Appendix}.
\section{Choosing Characteristic Timescales}
\label{C5:Sec:dSFR_Time-scales}
From the measured \acp{SFR} in Figure~\ref{C5:Fig:dSFR}, we can see that the techniques above drastically over-predict the intrinsic \ac{SFR} from the \acs{D14} \acs{SPH} simulations when compared on the same timescale. However, as mentioned before, the techniques \acs{SFR24}, \acs{SFR70} and \acs{SFRIR} were designed for whole galaxies or large regions of galaxies where the star-formation sites are averaged out. The assumed characteristic timescales are longer, \citep[e.\,g.~$\delta t_* = \SI{100}{\Myr}$:][]{KenniEvans:2012,Murphy:2011,Calzetti:2010,Rieke:2009}, since they are related to the life-time of high-mass stars. For the "observed" star-forming region presented in this work, the simulations stops after \SI{7}{\Myr}, which is assumed to be the time after the first supernova goes off. The supernova would suppress star formation within a timescale much shorter than \SI{100}{\Myr} and would cause the emission related to star formation to decay but to what extent remains unclear. 
\subsection{Averaging over Large Scales}
\label{C5:Sec:dSFR_averaging}
We will now test whether we can obtain a better agreement between star-formation laws and our simulation, when considering star formation over longer timescales and on larger spatial scales. Currently, however, there exist no galactic simulations which cover enough dynamic range to recover the star-forming regions in the required detail. For now, we need to rely on the \acs{D14} \acs{SPH} simulations. Therefore, to do this, we set up a toy model of a galaxy consisting of $N_{\textup{clusters}}$ star-forming regions at different evolutionary steps $t$.

We set the number of regions within the "galaxy" to $N_{\textup{clusters}}=\num{1.2e5}$ at different evolutionary steps randomly selected from the range $t\in[\num{0},\num{100}]\,\si{\Myr}$. We interpolate the luminosities $\nu L_\nu(\SI{24}{\microns})$, $\nu L_\nu(\SI{70}{\microns})$ and $L(\textup{IR})$ for every region $n$ and compute the sum of the different luminosities:
\begin{eqnarray}
   \nu L_{\nu}^{\textup{galaxy}}(\SI{24}{\microns}) &=&\sum\limits_{n=0}^{N_{\textup{clusters}}}\nu L_{\nu,n}(\SI{24}{\microns}) \\
   \nu L_{\nu}^{\textup{galaxy}}(\SI{70}{\microns}) &=&\sum\limits_{n=0}^{N_{\textup{clusters}}}\nu L_{\nu,n}(\SI{70}{\microns}) \\
   L^{\textup{galaxy}}(\textup{IR})                          &=&\sum\limits_{n=0}^{N_{\textup{clusters}}}L_n(\textup{IR})
\end{eqnarray}
Since the simulation we use stops after \SI{7}{\Myr}, so we need to make assumptions about what happens to the emission after the end of the simulation, which corresponds to the time when the first supernova goes off. We consider two limiting cases:
\begin{enumerate}
    \item \textbf{Sharp Stop}\\
    The supernova disrupts the region very quickly and the infrared emission is suppressed on a very short timescale.
    \item \textbf{No Stop}\\
    We assume that the emission stays constant after the supernova has gone off, until \SI{100}{\Myr}.
\end{enumerate}
If we assume that the emission stops sharply after the first supernova, we extract a lower limit on the total luminosity of: 
\begin{eqnarray}
   \nu L_{\nu}^{\textup{galaxy}}(\SI{24}{\microns})\big|_{min}&=&\SI{2.37e42}{\ergs\per\second}\\
   \nu L_{\nu}^{\textup{galaxy}}(\SI{70}{\microns})\big|_{min}&=&\SI{2.16e42}{\ergs\per\second}\\
   L^{\textup{galaxy}}(\textup{IR})                         \big|_{min}&=&\SI{4.61e42}{\ergs\per\second}.
\end{eqnarray}
However, if we assume that the emission will be constant after the first supernova, we recover an upper limit on the total luminosity of:
\begin{eqnarray}
   \nu L_{\nu}^{\textup{galaxy}}(\SI{24}{\microns})\big|_{max}&=&\SI{4.13e43}{\ergs\per\second}\\
   \nu L_{\nu}^{\textup{galaxy}}(\SI{70}{\microns})\big|_{max}&=&\SI{5.39e43}{\ergs\per\second}\\
   L^{\textup{galaxy}}(\textup{IR})                         \big|_{max}&=&\SI{7.28e43}{\ergs\per\second}.
\end{eqnarray}
Note that the upper as well as lower limit of the total luminosity lie within the range of validity of the \acs{SFR24} and \acs{SFR70} techniques. 

Following Section~\ref{C5:Sec:dSFR_SFR24}, Section~\ref{C5:Sec:dSFR_SFR70} and Section~\ref{C5:Sec:dSFR_SFRIR}, we calculate the rate $\textup{SFR}^{\textup{galaxy}}_\lambda(\delta t_*=\SI{100}{\Myr})$ of our toy "galaxy" for the lower limit of immediately suppressed emission after the supernova (case: sharp stop):
\begin{eqnarray}
    \label{C5:Eq:SFR_gal1}
   \textup{SFR}_{\SI{24}{\microns}}^{\textup{galaxy}}(\delta t_*=\SI{100}{\Myr})\big|_{min}&=&\SI{0.48}{\Msun\per\yr}\\
   \textup{SFR}_{\SI{70}{\microns}}^{\textup{galaxy}}(\delta t_*=\SI{100}{\Myr})\big|_{min}&=&\SI{0.13}{\Msun\per\yr}\\
   \textup{SFR}_{\textup{IR}}^{\textup{galaxy}}(\delta t_*=\SI{100}{\Myr})\big|_{min}               &=&\SI{0.18}{\Msun\per\yr}.
\end{eqnarray}
For the constant continuous emission after the first supernova (case: no stop) we calculate:
\begin{eqnarray}
   \textup{SFR}_{\SI{24}{\microns}}^{\textup{galaxy}}(\delta t_*=\SI{100}{\Myr})\big|_{max}&=&\SI{8.38}{\Msun\per\yr}\\
   \textup{SFR}_{\SI{70}{\microns}}^{\textup{galaxy}}(\delta t_*=\SI{100}{\Myr})\big|_{max}&=&\SI{3.17}{\Msun\per\yr}\\
   \label{C5:Eq:SFR_gal4}
   \textup{SFR}_{\textup{IR}}^{\textup{galaxy}}(\delta t_*=\SI{100}{\Myr})\big|_{max}               &=&\SI{2.83}{\Msun\per\yr}.
\end{eqnarray}

We now compute the intrinsic rate $\textup{SFR}^{\textup{galaxy}}_{\textup{sim}}$ of our toy "galaxy". We can estimate the total stellar mass of the "galaxy" $M_*^{\textup{galaxy}}$ by summing over the simulated stellar mass $M_{*,n}$ of every star-forming region at their assigned age in the "galaxy":
\begin{eqnarray}
   M_*^{\textup{galaxy}}&=& \sum\limits_{n=0}^{N_{\textup{clusters}}}M_{*,n}.
\end{eqnarray}
We found that $M_*^{\textup{galaxy}}=\SI{8.40e7}{\Msun}$ for the global timescale of $\delta t_* = \SI{100}{\Myr}$ and therefore estimate the \ac{SFR} averaged over \SI{100}{\Myr} to be:
\begin{eqnarray}
   \textup{SFR}^{\textup{galaxy}}_{\textup{sim}}(\delta t_*=\SI{100}{\Myr}) &=& \frac{M_*^{\textup{galaxy}}}{\delta t_*} = \SI{0.84}{\Msun\per\yr}.
\end{eqnarray}
We can compare the values of Eq.~\ref{C5:Eq:SFR_gal1} to Eq.~\ref{C5:Eq:SFR_gal4} with the actual rate of the "galaxy" $\textup{SFR}^{\textup{galaxy}}_{\textup{sim}}(\delta t_*=\SI{100}{\Myr})$ and find that the real value lies within the limits of the measurement from the \acs{SFR24}, \acs{SFR70}, and \acs{SFRIR} techniques. The measurement from the \acs{SFR24} technique, with a sharp emission cut-off, lies closest to the actual value. This is interesting because \ac{MIPS} \SI{24}{\microns} traces very recent emission, while the \ac{FIR} includes dust emission unrelated to star formation.

Note that the wide width of the limits is due to our lack of knowledge of how fast and with what function the emission from the star-formation sites will decay after the first supernova goes off. Most likely the real value is close to the sharp drop in emission. Nevertheless, we would like to emphasize that the fact that the measurements lie within the regions does not prove that the techniques \acs{SFR24}, \acs{SFR70} and \acs{SFRIR} work on the global scales for which they have been designed for, but rather show that the reason that the methods do not work on a single region is because the emission from the region needs to be averaged over longer timescales. Further tests from small scales to large scales are needed to verify these methods.
\begin{figure*}[t]
    \includegraphics[width=\textwidth]{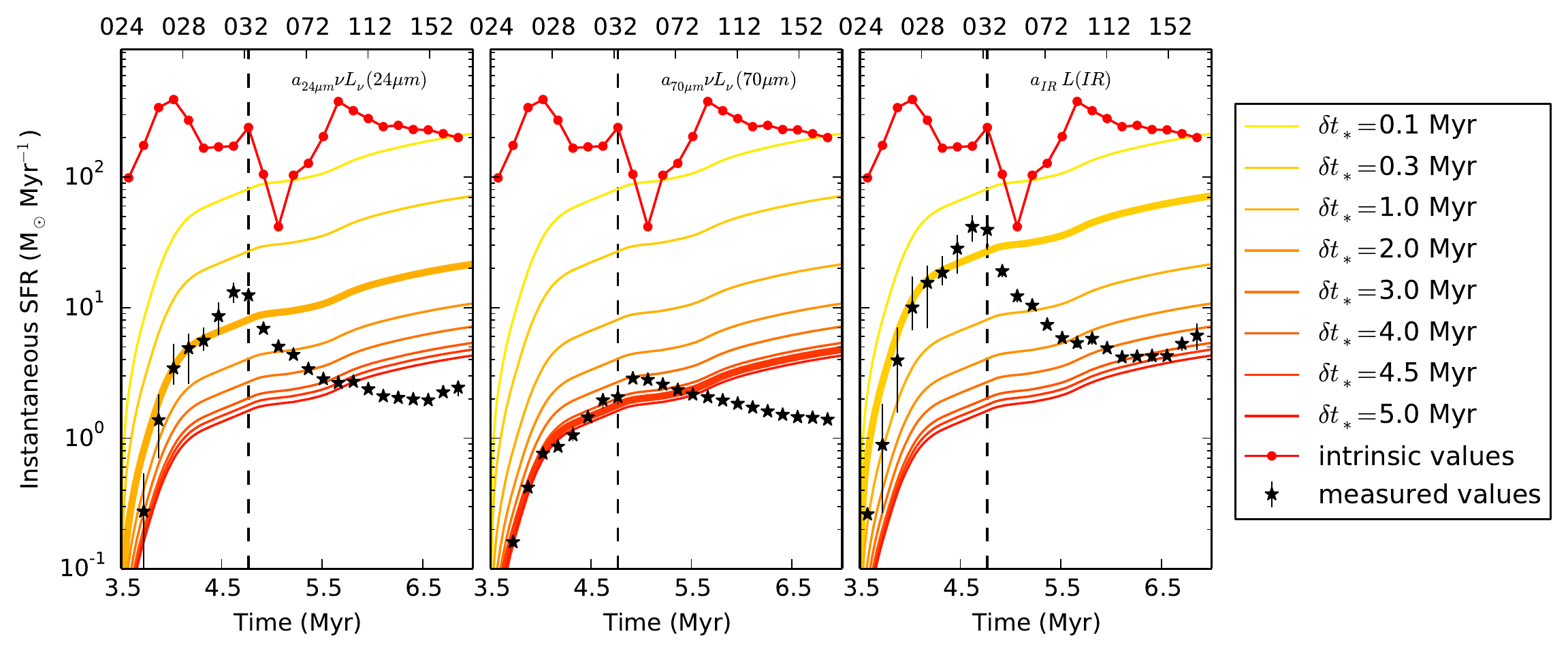}
  \caption[SFR rates convolved by different timescales]{\label{C5:Fig:dSFRconv}Instantaneous rate $\textup{SFR}_{\textup{sim}}(\delta t_* =\Delta t)$ from the simulation (red dots), rate convolved with different characteristic timescales $\delta t_*$ (yellow to red) and the scaled measured emission (black stars). Numbers on top represent the time-step IDs and vertical lines highlight the time-step of the simulations when feedback was switched on.}
\end{figure*}
%

\newpage
\subsection{Calibration for Small Scales}
\label{C5:Sec:dSFR_scaling}
From Section~\ref{C5:Sec:dSFR_SFR24} to Section~\ref{C5:Sec:dSFR_SFRIR}, we showed that the measured values $\textup{SFR}_{\SI{24}{\microns}}(\delta t_* =\SI{100}{\Myr})$, $\textup{SFR}_{\SI{70}{\microns}}(\delta t_* =\SI{100}{\Myr})$ and $\textup{SFR}_{\textup{IR}}(\delta t_* =\SI{100}{\Myr})$ from the techniques \acs{SFR24}, \acs{SFR70} and \acs{SFRIR}, respectively, cannot reproduce the actual rate $\hat{\textup{SFR}}_{\textup{sim}}(\delta t_* =\SI{100}{\Myr})$ of the simulation on a size-scale of \SI{30}{\pc}. In Section~\ref{C5:Sec:dSFR_averaging}, we showed however that for global scales which are governed by timescales of \SI{100}{\Myr}, the techniques work within the loose physical boundaries but require further investigation. In this section, we investigate whether the \acs{SFR24}, \acs{SFR70} and \acs{SFRIR} techniques could be improved for smaller scale star-forming regions.

In Figure~\ref{C5:Fig:dSFR}, we saw a mismatch of the measured properties (black) and simulated properties (red). The difference is due to two effects:
\begin{enumerate}
    \item The vertical scaling is off, which is due to the scaling factor $a$ from Eq.~\ref{SFRlaws}. This factor needs to be adjusted when measuring on smaller scales. 
    \item The shape of the measured and simulated function differs. A difference of shape is due to a difference of timescale. On the scale of this individual region, the \ac{MIPS} \SI{24}{\microns} emission appears to increase over \SI{1}{\Myr} before stabilizing due to the onset of high-mass stellar feedback.
\end{enumerate} 
\begin{figure*}[t]
    \includegraphics[width=\textwidth]{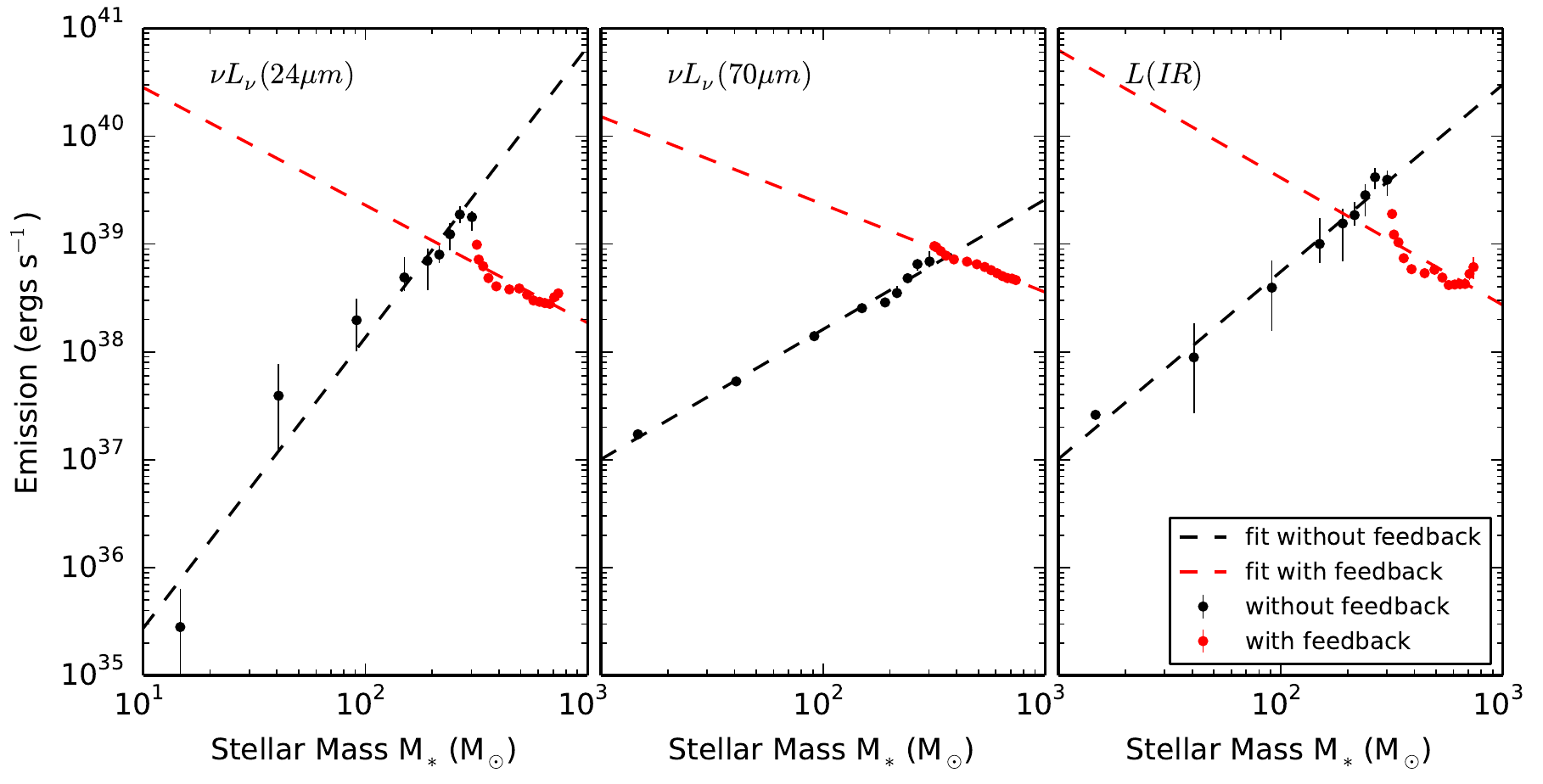}
  \caption[Relation of infrared emission and stellar mass in the simulated star-forming region]{\label{C5:Fig:dSFR_LM}Relation of infrared emission and stellar mass in the simulated star-forming region; (black) time-steps without high-mass stellar feedback acting; (red) high-mass stellar feedback acts on region.}
\end{figure*}
In order to explore how we might achieve a better agreement, we again \citep[see][]{Bronstein,NumericalRecipes} convolve the intrinsic rate $\textup{SFR}_{\textup{sim}}(\delta t_* =\Delta t)$ with normalized top-hat functions of timescales $\delta t_*$, reaching from \SIrange{0.1}{5.0}{\Myr}, to reproduce the shape of the measured emission. We then vary the scaling factor $a_\lambda$ for every measurement technique until a good fit between the convolved rate $\hat{\textup{SFR}}_{\textup{sim}}(\delta t_*)$, at the new timescale $\delta t_*$, and the measured emission is reached.

In Figure~\ref{C5:Fig:dSFRconv}, we show the $\textup{SFR}_{\textup{sim}}(\delta t_* =\Delta t)$ (red dots) from the simulation and the convolved counterparts for different characteristic timescales $\delta t_*$. We highlighted the best fit of the convolved function $\hat{\textup{SFR}}_{\textup{sim}}(\delta t_*)$ and the scaled emission $a_{\SI{24}{\microns}} \nu L_\nu(\SI{24}{\microns})$, $a_{\SI{70}{\microns}} \nu L_\nu(\SI{70}{\microns})$ or $a_{\textup{IR}} \nu L(\textup{IR})$ as a thick line. Our fitted parameters for $\delta t_*$ and $a_\lambda$ are:
\begin{eqnarray}
  a_{\SI{24}{\microns}} = \num{7e-45} && \delta t_* \approx \SI{1.0}{\Myr}\\
  a_{\SI{70}{\microns}} = \num{3e-45} && \delta t_* \approx \SI{4.5}{\Myr}\\
  a_{\textup{IR}} = \num{1e-44} && \delta t_* \approx \SI{0.3}{\Myr}.
\end{eqnarray}

Nevertheless, when inspecting Figure~\ref{C5:Fig:dSFRconv}, one can see that it is challenging to pin-down a characteristic timescale, since for simulation time-steps where high-mass stellar feedback is present, the region is governed by different physical processes which also suppress the emission at later times.

\section{Evolution of the Emission}
\label{C5:Sec:dSFRemission}
%
As noted in Section~\ref{C5:Sec:dSFR_Time-scales}, we "observe" a drop in emission once the high-mass stellar feedback is switched-on. In Figure~\ref{C5:Fig:dSFR_LM}, we plot the evolution of the emission $\nu L_\nu (\SI{24}{\microns})$, $\nu L_\nu (\SI{70}{\microns})$ and the total infrared luminosity $L (\textup{IR})$ versus the stellar mass $M_*$. Since the points are equally spaced in time, we can see that the infrared emission increases steeply with increasing stellar mass in very few time-steps before (black) the stars start to ionize. Afterwards (red), the emission actually goes down with slightly increasing mass over many time-steps. We can see this evolution equally for the two monochromatic emissions in the \ac{MIR} and \ac{FIR} and also for the total infrared emission. We tried to quantify the change in emission of the two different episodes through a power-law fit of the following function:
\begin{eqnarray}
  \frac{L}{\si{\ergs\per\second}} &=& b \left(\frac{M_*^{\textup{sim}}}{\si{\Msun}}\right)^c
\end{eqnarray}
We found from the "observed" set of synthetic observations that for regions where no high-mass stellar feedback is present, the emission versus stellar mass relation is represented very well by a power-law. The parameters for the different luminosities are:
\begin{eqnarray}
  b_{\SI{24}{\microns}} = \num{5.5e32} && c_{\SI{24}{\microns}} = \num{2.70} \\
  b_{\SI{70}{\microns}} = \num{6.3e35} && c_{\SI{70}{\microns}} = \num{1.21} \\
  b_{\textup{IR}}                = \num{1.9e35} && \ \ \,c_{\textup{IR}}          = \num{1.74}.               
\end{eqnarray}
When high-mass stellar feedback is present, the emission versus stellar mass relation also well fit by a power-law with the following fitting parameters:
\begin{eqnarray}
  b_{\SI{24}{\microns}} = \num{3.5e41} && c_{\SI{24}{\microns}} = \num{-1.09} \\
  b_{\SI{70}{\microns}} = \num{9.9e40} && c_{\SI{70}{\microns}} = \num{-0.81} \\
  b_{\textup{IR}}                = \num{9.5e41} && \ \ \,c_{\textup{IR}}          = \num{-1.18}.                
\end{eqnarray}
These relations indeed show a very interesting development once the high-mass stellar feedback is acting in the star-forming region. The tilt in the power-law of the two different episodes can be explained through the ionizing bubble which is driven by the high-mass stellar feedback: Without high-mass stellar feedback most of the dust emitting in the \ac{MIR} is located closer to the \acsp{YSO} than the emission from the \ac{FIR}. Once the stellar objects ionize and drive winds, the circumstellar material from this stars is eroded. Close-by accreting stars also lose their outer circumstellar material. Therefore, the turn-over of the power-laws is very rapid and cleaner for the \SI{24}{\microns} emission than for the \SI{70}{\microns} emission, as can be seen in Figure~\ref{C5:Fig:dSFR_LM} (left \& middle). Subsequently, the closer circumstellar material is eroded and the power-law in the Figure~\ref{C5:Fig:dSFR_LM} tilts to negative values. The decreasing gas mass within the field of view also has an effect on the decreasing emission, however, this is not the dominating process. The same happens for the total infrared emission which is just a combination of the monochromatic \SI{24}{\microns} and \SI{70}{\microns} emission. 

Note that even while high-mass stellar feedback is present in star-forming regions, star formation by itself does not stop completely. At the rim of the ionizing bubble, stars are still forming and growing.

\section{Proto-Stellar Counting}
\label{C5:Sec:cSFR}
\begin{figure*}[t]
\includegraphics[width=\textwidth]{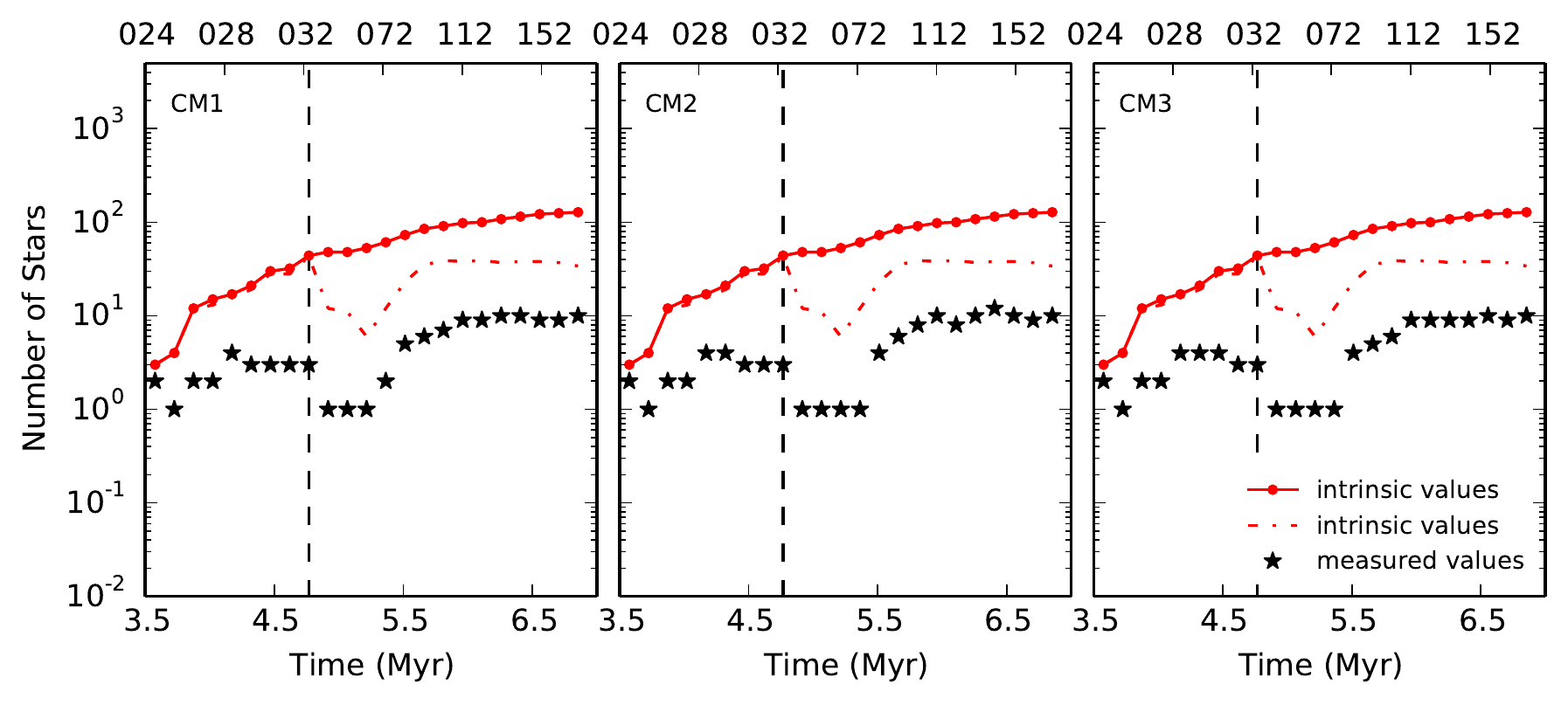}
\caption[Counted number of stars for the different circumstellar set-ups]{\label{C5:Fig:cN_star} Counted number of stars for the different circumstellar set-ups (black) in comparison to the actual number of stars (red dots) and the actual number of accreting stars (dashed).}
\end{figure*}

We analyzed the point sources in our synthetic star-forming region and produced a point-source catalogue. In a follow-up paper \citepalias{KDR3:inprep}, we will perform a detailed analysis regarding the \ac{SFR}. Here, we concentrate on studying the point sources in the different circumstellar set-ups. In Figure~\ref{C5:Fig:cN_star}, we present the number of point sources for the different time-steps and circumstellar set-ups.

In Section~\ref{C5:Sec:dSFRemission}, we found that the number of accreting stars in the region is less than the number of young stars because some former accreting low-mass stars lost their material due to the high-mass stellar feedback of the neighboring stars. In Figure~\ref{C5:Fig:cN_star}, the red dots represent the total number of stars (hence, ionizing stars, eroded \acsp{YSO} and accreting \acsp{YSO}), while the dashed line represents the number of accreting \acsp{YSO}. We can see that the counted point sources (black stars) follow the trend of the accreting \acsp{YSO} (dashed line). From Figure~\ref{C5:Fig:cN_star} (red dashed line), we can see that the number of accreting stars declines after the high-mass stellar feedback is switched on, but stabilizes again. The dip in the relation can be explained be cause the onset of high-mass stellar feedback redistributes star-formation in space and time.

The trend of the measured point sources is similar for the different circumstellar set-ups. However, we found that the set-up without circumstellar refinement (\acs{CM1}) produced the brightest point sources. The brightness of forming stars with rotationally flattened envelope profiles (\acs{CM2}) and with extrapolated power-law profiles (\acs{CM3}) produces similar results. However, in some cases the accreting \acsp{YSO} in \acs{CM3} remained undetected due to their very steep power-law profiles at the center of the envelopes.

Nevertheless, the number of accreting stars is about four times higher than the counted point sources, regardless the circumstellar set-up. This is mostly due to the detection limit in the synthetic star-forming region. In a follow-up paper \citepalias{KDR3:inprep}, we will perform a population synthesis study to estimate the \ac{SFR} directly from the young stellar population. The corresponding assumption of the \ac{IMF} and the timescales will be studied in detail.

\section{Discussion}
\label{C5:Sec:Discussion}
In the previous sections, we tested techniques commonly applied by observers to measure star-formation properties. We will now discuss the shortcomings and challenges when testing and measuring these properties:

\subsection{Choice of Diffuse Star-formation Tracers}
The \ac{SFR} measurement techniques from Section~\ref{C5:Sec:dSFR} hold under the assumption that the infrared flux is solely due to the absorbed and re-emitted \ac{UV} radiation from young high-mass stars. However, not all infrared emission, especially the \ac{FIR}, is due to the channelling of \ac{UV} photons. Hydrodynamical heating of the medium and the radiative heating of low-mass stars can also contribute to the flux. The feedback of low-mass young stars is important, since most of the stars are of low mass. 

However, as stated before, the \SI{24}{\microns} emission mostly originates from star-formation sites and can act as a better proxy than the \ac{FIR}. Techniques that make use of the total emission beyond \SI{70}{\microns}, such as \acs{SFR70} and \acs{SFRIR}, are difficult to interpret, since they are simultaneously tracers of dust and gas (c.\,f.~\citetalias{KDR2a:inprep}), rather than just tracers of sites of actual star formation and, therefore, the emission is also affected by the processes (e.\,g.~high-mass stellar feedback) away from the individual star-formation sites (c.\,f.~Section~\ref{C5:Sec:dSFRemission}).

\subsection{Biases}
\label{C5:Sec:Discussion_biases}
In what follows, we describe the biases introduced when measuring star-formation properties:

\subsubsection{Reliability of Local and Global Characteristic Time-scales}
When we observe local star formation on small scales, we probe regions that are governed by much shorter timescales, while for global scales (such as galaxies), variations of star formation happen over much longer timescales \citep{Kruijssen:2014}. In Section~\ref{C5:Sec:dSFR_Time-scales}, we explored the effect different timescales have on the recovered rates. We found that it is hard to single out one characteristic timescale which represents all the episodes of a star-forming region which are governed by different physics (e.\,g.~early phases without feedback, high-mass stellar feedback, after first supernova). 

Of course, these effects are averaged out when we observe galaxies, and one has to keep in mind that the recovered \ac{SFR} is not the actual instantaneous \ac{SFR} in the regions but just an average over the entire observation. In future, to test the accuracy of star-formation tracers on galaxy scales, one would of course ideally require a global simulation of a galaxy that also resolves individual forming stars. 

\subsubsection{Reliability of the Measured Star-formation Rates}
In Section~\ref{C5:Sec:dSFR}, we showed that empirical global diffuse star-formation tracers cannot be applied to local, small-scale regions. We showed examples of the calibration of \acp{SFR} for single star-forming regions by introducing new characteristic timescales and scaling factors. We note that for the small scales, where we probe the actual physics of star formation and where different processes act at different times (such as high-mass stellar feedback), it is difficult to single out one characteristic timescale and scaling factor for the entire star-forming region. 

If our new relations to measure the \ac{SFR} are used anyway, then we suggest the use of diffuse tracers only for regions at early stages, where ionization and winds are not present. From our synthetic observations combined with a background, we found that it is easiest to disentangle the star-forming region at \SI{24}{\microns} from the background of the Galactic plane (compared to longer wavelengths). Therefore, biases due to the complex background are minimized. 

\subsubsection{Propagation of Biases}
We would like to note that since there are still large biases when measuring the \ac{SFR} from diffuse tracers. Further, the extracted stellar masses of a region are also biased, since they are a product of the \ac{SFR} and characteristic timescale $\delta t_*$ which are both very uncertain. In \citepalias{KDR2a:inprep}, we showed that the estimated total gas masses of an individual star-forming region using modified blackbody fitting can produce reliable results in moderate background regions. However, the from the technique independent systematic errors (e.\,g.~distance estimates) bias the total gas mass. Naturally, the estimates for the \ac{SFE} (Eq.~\ref{C5:Eq:SFE}) which is just a combination of the stellar mass and the total gas mass should therefore be treated with caution.

\subsection{Impact of Feedback}
\label{C5:Sec:Discussion_feedback}
In Section~\ref{C5:Sec:dSFRemission}, we showed that high-mass stellar feedback (ionization \& winds) have a strong effect on the emission of the cloud in the infrared. The growing ionization bubble removes circumstellar material around the high-mass stars, which can be first observed at \ac{MIR} wavelengths and then at later time-steps for \ac{FIR} emission. The circumstellar envelopes of low-mass accreting stars are also eroded from the outside.

The simulations of \acs{D14} could be used further to explore the different feedback mechanisms on the synthetic observations, since they extensively modeled different combinations of ionization and winds. A finer grid of time-steps would be preferable to study the timescale under which the different feedback scenarios affect the emission in the different bands. Furthermore, simulations which include feedback physics on earlier time-steps in the simulation would be preferable. Currently in the \acs{D14}, high-mass stellar feedback is switched on only after three stars have reached or exceeded the stellar mass of \SI{20}{\Msun} and their accretion is suppressed. However, in reality, high-mass stars will ionize their surrounding in a continuous process while growing mass and the ionization processes is not "switched on" in a sharp manner.

\subsection{Tracer for Stellar Mass}
We found that on small scales, the infrared emission is a much better tracer of stellar mass $M_*$ than of the \ac{SFR}.

\subsubsection{Early Phases}
At early times in the simulations, where there is no high-mass stellar feedback present, we showed (see Section~\ref{C5:Sec:dSFRemission}) that it is possible to define a simple relation $\frac{L}{\si{\ergs\per\second}} = b \left(\frac{M_*}{\si{\Msun}}\right)^c$ between the infrared luminosities (at \SI{24}{\microns}, \SI{70}{\microns} or total infrared) and the stellar mass $M_*$ of the region. We suggest using \SI{24}{\microns} emission to relate to the total stellar mass $M_*$ in the star-forming region because the background is less complex in the \ac{MIR} and the flux can be corrected more easily. Of course, this relation needs to be tested using other simulations before it can be generalized into a law that can be used on observed star-forming regions.

\subsubsection{High-Mass Stellar Feedback Phases}
Once the high-mass feedback is turned on, the infrared emission decreases, because the infrared emission is also a tracer of the dust. Changes in the geometry of the region, due to the high-mass stellar feedback, for example, pushing the dust further from the stars or destroying it \citep{DiazMiller:1998, Mathews:1967}, can result in a decrease in the infrared emission. 

As a result, we recommend against using the infrared luminosity as an accurate tracer of star formation on small scales where high-mass stellar feedback, such as ionization and winds, dominate.

\subsection{Stellar Population}
\label{C5:Sec:Discussion_pop}
Here, we will discuss shortcomings and advances when sampling the stellar population directly:

\subsubsection{Circumstellar Set-up}
In \citetalias{KDR1:inprep}, we described different circumstellar set-ups. In this paper (referred to as \citetalias{KDR2b:inprep}), we explored the synthetic observations resulting from \citetalias{KDR1:inprep} and we showed that the different set-ups do not affect the total flux of the region at \SI{24}{\microns} much. However, the detectability of the point sources is dependent on the different types of inward extrapolation of the envelope profile. 

We can see that numbers of measured point sources are similar for the different circumstellar set-ups. However, we found that the set-up without circumstellar refinement (\acs{CM1}), produces the brightest point sources. The brightening for the objects without envelopes is due to the larger volume of dust exposed to the stellar radiation at the resolution radius of the simulation and also the larger stellar luminosity since the stellar mass was set to the mass of the sink particle in this case (see \citealt{Koepferl:2015} and \citetalias{KDR1:inprep} for more details). The brightness of rotationally flattened profiles (\acs{CM2}) and the extrapolated power-law profiles (\acs{CM3}) produces similar results. However, in some cases the accreting \acsp{YSO} with the \acs{CM3} set-up remained undetected due to their very steep power-law profiles at the center. When synthetic observations are used to study point sources, we therefore suggest using the rotationally flattened envelope profile to refine beyond the resolution radius, since it is consistent with the physical description of an in-falling envelope. Nevertheless, we need to keep in mind that we deal here with low-number statistics.

\subsubsection{Initial Mass Function}
While the stellar mass is not biased due to characteristic timescales used when measuring the \ac{SFR} through proto-stellar counting, the estimate of the total stellar mass is dependent on the assumption taken about the mass function of the stellar population. Especially, when we are limited in sensitivity. It is currently under debate whether the initial mass functions \citep[e.\,g.~][]{Kroupa:2001, Salpeter:1955} hold for all local regions as well as on global scales (hence, in other galaxies) (see \citealt{Offner:2014}). \cite{Jappsen:2005} showed that the \ac{IMF} depends on the thermodynamical state of the star-forming gas. However, the star-forming gas might vary at different locations, due to temperature, pressure, metallicity and chemistry in general. 
Therefore, it is challenging to transfer the counted sample of \acsp{YSO} to a reliable \ac{SFR}. As an example, in the \acs{D14} simulations, during the time-steps where we see high-mass stellar feedback, and where we already have more than \num{100} stars, providing good number statistics, the simulation produces time-step dependent top heavy \ac{IMF}. In the follow-up paper \citepalias{KDR3:inprep}, we will explore the effects of the chosen \ac{IMF} in more detail.

\subsubsection{Diffuse Star-formation Proxies versus Direct Counting}
In this paper \citepalias{KDR2b:inprep}, we studied indirect diffuse star-formation tracers and showed that the commonly applied star-formation laws produce values that are too high when the techniques are applied to single star-forming regions (see Section~\ref{C5:Sec:dSFR}). We showed that the techniques can be scaled in order to use them for single star-forming regions, even though the actual shape and pace of the instantaneous \ac{SFR} is not recovered. However, when inspecting the detected point sources from our point-source catalogue (see Section~\ref{C5:Sec:cSFR}), we found that the number of detected point sources follows the same trend as the number of accreting objects over time and is comparable to the instantaneous \acs{SFR}.
\section{Summary}
\label{C5:Sec:Summary}
In this paper \citepalias{KDR2b:inprep}, we explored different measurement techniques of star-formation properties such as the $\textup{SFR}_{\SI{24}{\microns}}$, $\textup{SFR}_{\SI{70}{\microns}}$, $\textup{SFR}_{\textup{IR}}$ and the gas mass $M_{\textup{gas}}$ from modified blackbody fitting for $\sim$5800 synthetic observations of a $\sim$\,\SI{30}{\pc} star-forming region at different evolutionary time-steps, orientations, distances and different circumstellar radiative transfer set-ups. We further explored the accuracy of the techniques and the implications for small and for global scales:
\begin{itemize}
\item {\bf Diffuse Monochromatic and Total Infrared Star-formation Tracers}\\
On our synthetic star-forming region, we tested the diffuse monochromatic star-formation tracers of the total \SI{24}{\microns} and \SI{70}{\microns} emission originally designed for more global scales (\citealt{Rieke:2009} and \citealt{Calzetti:2010}, respectively). Further, we explored the accuracy of the total infrared tracers \citep{Murphy:2011}. We showed that these star-formation tracers deviate by several orders of magnitude from the intrinsic value. Further, we showed that in the \ac{FIR}, it becomes more and more difficult to disentangle the synthetic region from the non-uniform background in the Galactic plane. We conclude that the star-formation tracers, using the diffuse infrared tracers developed for the extra-galactic community, do not work for our single star-forming region and hence, do not work on local scales, because we observe the peak of star formation and not its average, as this is the case for larger scales. Therefore, we suggest not to use these tracers as a star-formation proxy of single star-forming regions.
\item {\bf Characteristic Time-scales for Star-forming Regions}\\
We explored which characteristic timescale relates the observed emission in the infrared with the actual star formation. We singled out timescales for regions not dominated by high-mass stellar feedback of \SIlist{1.0;4.5;0.3}{\Myr} for the \SI{24}{\microns}, \SI{70}{\microns} and the total infrared emission, respectively. We further adjusted the star-formation laws (see Eq.~\ref{SFRlaws}) for single star-forming regions, which do not yet experience high mass stellar feedback. We found scaling factors of \numlist{7e-45;3e-45;1e-44} for the \SI{24}{\microns}, \SI{70}{\microns} and the total infrared emission, respectively. However, we suggest using the \SI{24}{\microns} star-formation tracer because it traces the direct sites of star formation and is not easily contaminated by Galactic emission (e.\,g.~from the Galactic plane), which otherwise would lead to large biases. 
\item {\bf Global Diffuse Star-formation Tracers}\\
We further showed that for objects of larger scales, such as parts of galaxies, the diffuse star-formation tracers produce more consistent results. However, further tests, preferably with simulations with a larger dynamic range, are needed to verify the accuracy of the diffuse star-formation tracers on global scales and hence, for extra-galactic applications.
\item {\bf Impact of Feedback on the Emission}\\
We found that high-mass stellar feedback affects the infrared emission almost instantly and produces a tip of the slope in the stellar mass versus infrared luminosity relation from positive to negative slopes. We found that since high-mass stellar feedback erodes the material first from inner regions of the \acsp{YSO}, the emission at \SI{24}{\microns} is affected strongly compared to the \SI{70}{\microns}. Further tests with regions of different wind and different ionization power are needed to explore these effects further. 
\item {\bf Tracer for Stellar Mass}\\
We found that infrared emission is a much better tracer of stellar mass $M_*$ than of the \ac{SFR} on small scales and for phases without high-mass stellar feedback. We determine a power-law relation $\frac{L_{\SI{24}{\microns}}}{\si{\ergs\per\second}} = \num{5.5e32} \left(\frac{M_*}{\si{\Msun}}\right)^{\num{2.70}}$ between the \SI{24}{\microns} emission and the total stellar mass $M_*$. However, this relation breaks down once the stars start to ionize and drive winds.
\item {\bf Circumstellar Radiative Transfer Set-up}\\
We explored the use of point sources at \SI{24}{\microns} to derive the \ac{SFR}. While the different circumstellar set-ups do not affect the total flux at \SI{24}{\microns} much, the detectability of the point sources is dependent on the different types of inward extrapolation of the envelope profile. We found that the set-up without inwards extrapolation produces brighter point sources than the set-ups with the rotationally flattened and power-law envelopes. When synthetic observations are used to study point sources, we therefore suggest using the  rotationally flattened envelope profile to refine beyond the resolution radius, since it is consistent with the physical description of an in-falling envelope. 
\item {\bf Proto-stellar Counting}\\
We explored the proto-stellar counting technique at \SI{24}{\microns} and recovered a point-source catalogue for the different circumstellar set-ups. We showed that the trend of the detected point sources over time follows the trend of the number of accreted objects and hence the instantaneous \ac{SFR}. In a follow-up paper \citepalias{KDR3:inprep}, we will perform a detailed study of the recovered \acp{SFR} from proto-stellar counting.
\end{itemize}

We provide our measurements of the \ac{SFR} from the different techniques and the measured gas mass in Table~\ref{C5:Appendix:data_CMxOxDx} of Appendix~\ref{C5:Appendix}. Moreover, the dust temperature and dust surface density maps will be provided in the online material of Appendix~\ref{C5:Appendix}.

\section{Acknowledgements}
We thank the referee for a constructive report that helped us improve the clarity and the strength of the results presented in our paper. This work was carried out in the Max Planck Research Group \textit{Star formation throughout the Milky Way Galaxy} at the Max Planck Institute for Astronomy. C.K. is a fellow of the International Max Planck Research School for Astronomy and Cosmic Physics (IMPRS) at the University of Heidelberg, Germany and acknowledges support. C. K. acknowledges support from STFC grant ST/M001296/1.
J. E. D. was supported by the DFG cluster of excellence \textit{Origin and Structure of the Universe}. 
This research made use of Astropy, a community-developed core Python package for Astronomy \citep{Astropy:2013}, matplotlib, a Python plotting library \citep{Hunter:2007}, Scipy, an open source scientific computing tool \citep{Scipy}, the NumPy package \citep{NumPy} and IPython, an interactive Python application \citep{IPython}.

\bibliographystyle{latex_apj}
\bibliography{latex_ref}

\appendix

\section{Online Material for the Analysis of Synthetic Star-forming Regions}
\label{C5:Appendix}

In this appendix, we list the measured and simulated star-formation rates for the different combinations of circumstellar set-ups (\acs{CM1}: no refinement, \acs{CM2}: refinement with rotationally flattened envelope, \acs{CM3}: refinement with power-law envelope), orientations (\acs{O1}: xy plane, \acs{O2}: xz plane, \acs{O3}: yz plane) and distances (\acs{D1}: \SI{3}{\kpc}, \acs{D2}: \SI{10}{\kpc}).

\begin{longtable}{ccccrccccccc}
\caption[Measured Star-formation Rates]{\label{C5:Appendix:data_CMxOxDx}Measured Star-formation Rates for different time-steps and setups}\\
\hline\\[-13pt]
\hline\\[-5pt]
&&&&&&&&&&\\
setup & run & \multicolumn{1}{c}{time} & \multicolumn{1}{c}{$\hat{\mbox{SFR}}_{\textup{sim}}^{\delta t_* =\SI{100}{\Myr}}$} &
 \multicolumn{1}{c}{$\mbox{SFR}_{\SI{24}{\microns}}^{\textup{B1}}$}& \multicolumn{1}{c}{$\mbox{SFR}_{\SI{24}{\microns}}^{\textup{B2}}$} &
 \multicolumn{1}{c}{$\mbox{SFR}_{\SI{70}{\microns}}^{\textup{B1}}$}& \multicolumn{1}{c}{$\mbox{SFR}_{\SI{70}{\microns}}^{\textup{B2}}$} &
 \multicolumn{1}{c}{$\mbox{SFR}_{\textup{IR}}^{\textup{B1}}$}& \multicolumn{1}{c}{$\mbox{SFR}_{\textup{IR}}^{\textup{B2}}$}\\
(IDs) & (ID) & \multicolumn{1}{c}{(\si{\Myr})} & (\si{\Minfall}) & (\si{\Minfall}) & (\si{\Minfall}) & (\si{\Minfall}) & (\si{\Minfall}) & (\si{\Minfall}) & (\si{\Minfall}) \\[6pt]
 				\hline\\[-5pt]
\endfirsthead
\endhead
CM1O1D1 & 024 & 3.576 & \num{2.24e-09} & \num{2.51e-08} & \num{3.48e-05} & \num{1.01e-06} & \num{4.33e-05} & \num{1.00e-06} & \num{7.34e-05} \\
CM1O1D1 & 025 & 3.725 & \num{8.08e-09} & \num{3.99e-06} & \num{3.87e-05} & \num{3.06e-06} & \num{4.53e-05} & \num{1.69e-06} & \num{7.40e-05} \\
CM1O1D1 & 026 & 3.874 & \num{1.91e-08} & \num{3.13e-05} & \num{6.61e-05} & \num{8.00e-06} & \num{5.03e-05} & \num{1.15e-05} & \num{8.37e-05} \\
CM1O1D1 & 027 & 4.023 & \num{3.47e-08} & \num{1.53e-04} & \num{1.87e-04} & \num{1.45e-05} & \num{5.68e-05} & \num{6.69e-05} & \num{1.39e-04} \\
CM1O1D1 & 028 & 4.172 & \num{4.88e-08} & \num{8.15e-05} & \num{1.16e-04} & \num{1.57e-05} & \num{5.79e-05} & \num{3.08e-05} & \num{1.03e-04} \\
CM1O1D1 & 029 & 4.321 & \num{5.80e-08} & \num{1.93e-04} & \num{2.27e-04} & \num{2.41e-05} & \num{6.64e-05} & \num{9.18e-05} & \num{1.63e-04} \\
CM1O1D1 & 030 & 4.470 & \num{6.52e-08} & \num{2.70e-04} & \num{3.05e-04} & \num{2.80e-05} & \num{7.02e-05} & \num{1.32e-04} & \num{2.03e-04} \\
CM1O1D1 & 031 & 4.619 & \num{7.25e-08} & \num{3.89e-04} & \num{4.23e-04} & \num{3.35e-05} & \num{7.58e-05} & \num{1.74e-04} & \num{2.46e-04} \\
CM1O1D1 & 032 & 4.768 & \num{8.13e-08} & \num{3.96e-04} & \num{4.30e-04} & \num{3.67e-05} & \num{7.90e-05} & \num{1.83e-04} & \num{2.54e-04} \\
... & ... & ... & ... & ... & ... & ... & ... & ... & ... \\[6pt]
 				\hline\\[-5pt]
\end{longtable}
\vspace{-0.5cm}
The full table will be provided in the online material.

\begin{acronym}[ATLASGAL]
\acro{2MASS}{Two Micron All-Sky Survey}
\acro{AGB}{Asymptotic Giant Branch}
\acro{ALMA}{Atacama Large Millimeter/Submillimeter Array}
\acro{AMR}{Adaptive Mesh Refinement}
\acro{ATLASGAL}{APEX Telescope Large Area Survey of the Galaxy}
\acro{BGPS}{Bolocam Galactic Plane Survey}
\acro{c1}{sample clipping of only neutral particles within a box of \SI{30}{\pc}}
\acro{c2}{sample clipping of \acs{c1} particles and for a certain threshold temperature}
\acro{c2d}{Cores to Disks Legacy}
\acro{CASA}{Common Astronomy Software Applications package}
\acro{cm}{centimeter}
\acro{CMF}{core mass function}
\acro{CMZ}{central molecular zone}
\acro{D1}{distance at \SI{3}{\kpc}}
\acro{D14}{\acs{SPH} simulations performed by Jim Dale and collaborators \citep{DaleI:2011,DaleIoni:2012,DaleIoni:2013,DaleWind:2013,DaleBoth:2014}}
\acro{D2}{distance at \SI{10}{\kpc}}
\acro{DT1}{temperature coupling of radiative transfer \& hydrodynamical temperature}
\acro{DT2}{temperature coupling of the radiative transfer \& isothermal temperature}
\acro{DT3}{no temperature coupling of the radiative transfer temperature}
\acro{e1}{using the \cite{Ulrich:1976} envelope profile to extrapolate the envelope inwards}
\acro{e2}{using the \cite{Ulrich:1976} envelope profile with suppressed singularity to extrapolate the envelope inwards}
\acro{e3}{using a power-law envelope profile to extrapolate the envelope inwards}
\acro{EOS}{equation of state}
\acro{FIR}{far-infrared}
\acro{FITS}{Flexible Image Transport System}
\acro{FWHM}{full-width at half-maximum}
\acro{GLIMPSE}{Galactic Legacy Infrared Mid-Plane Survey Extraordinaire}
\acro{GMC}{Giant Molecular Clouds}
\acro{Hi-GAL}{\emph{Herschel} Infrared Galactic Plane Survey}
\acro{HST}{\emph{Hubble} Space Telescope}
\acro{HWHM}{half-width at half-maximum}
\acro{IMF}{initial mass function}
\acro{IR}{infrared}
\acro{IRAC}{Infrared Array Camera}
\acro{IRAS}{Infrared Astronomical Satellite}
\acro{ISM}{interstellar medium}
\acro{$K$}{K band}
\acro{LTE}{local thermodynamical equilibrium}
\acro{MIPS}{Multiband Imaging Photometer for \emph{Spitzer}}
\acro{MIPSGAL}{\acs{MIPS} Galactic Plane Survey}
\acro{MIR}{mid-infrared}
\acro{MS}{main-sequence}
\acro{mm}{millimeter}
\acro{NASA}{National Aeronautics and Space Administration}
\acro{NIR}{near-infrared}
\acro{N-PDF}{column density \acs{PDF}}
 \acro{O1}{xy plane}
 \acro{O2}{xz plane}
 \acro{O3}{yz plane}
  
\acro{p1}{parameter evaluation version from \acs{SPH} kernel function}
\acro{p2}{parameter evaluation version from \acs{SPH} splitted kernel distribution}
\acro{p3}{parameter evaluation version from \acs{SPH} random distribution}
\acro{PACS}{Photoconductor Array Camera and Spectrometer}
\acro{PAH}{polycyclic aromatic hydrocarbon}
\acro{PDF}{probability distribution function}
\acro{PDR}{Photon Dominated Region}
\acro{PSF}{point-spread-function}
\acro{px}{one of the parameter evaluation version \acs{p1}, \acs{p2}, \acs{p3}}
\acro{RGB}{red, green and blue}
\acro{CM1}{circumstellar setup with background density and sink mass as stellar mass}
\acro{CM2}{circumstellar setup by a toy model with \acs{e2} envelope superposition density and corrected stellar mass, protoplanetary disk and envelope cavity}
\acro{CM3}{circumstellar setup by a toy model with \acs{e3} envelope superposition density and corrected stellar mass, protoplanetary disk and envelope cavity}
\acro{s1}{Voronoi site placement version at \acs{SPH} particle position}
\acro{s2}{Voronoi site placement version as \acs{s1} including sites at sink particles}
\acro{s3}{Voronoi site placement version as \acs{s2} including circumstellar sites}
\acro{SAO}{Smithsonian Astrophysical Observatory}
\acro{SED}{spectral energy distribution}
\acro{SFE}{star-formation efficiency}
\acro{SFR}{star-formation rate}
\acro{SFR24}{technique to measure the \acs{SFR} using the \SI{24}{\microns} tracer}
\acro{SFR70}{technique to measure the \acs{SFR} using the \SI{70}{\microns} tracer}
\acro{SFRIR}{technique to measure the \acs{SFR} using the total infrared tracer}
\acro{Sgr}{Sagittarius}
\acro{SMA}{Submillimeter Array}
\acro{SPH}{smoothed particle hydrodynamics}
\acro{SPIRE}{Spectral and Photometric Imaging Receiver}
\acro{sub-mm}{sub-millimeter}
\acro{UKIDSS}{UKIRT Infrared Deep-Sky Survey}
\acro{UKIRT}{UK Infrared Telescope}
\acro{UV}{ultra-violet}
\acro{WFCAM}{\acs{UKIRT} Wide Field Camera}
\acro{WISE}{Wide-field Infrared Survey Explorer}
\acro{YSO}{young stellar object}
\end{acronym}

\end{document}